\documentclass[italian,english]{article}
\usepackage[latin9]{inputenc}
\usepackage{color}
\usepackage{amssymb}

\newcommand{\lyxaddress}[1]{
\par {\raggedright #1
\vspace{1.4em}
\noindent\par}
}

\usepackage{babel}

\begin{document}

\title{\textbf{Gravitational Waves Astronomy: a cornerstone for gravitational
theories}}

\author{\textbf{Christian Corda }}

\maketitle

\lyxaddress{\begin{center}
Associazione Scientifica Galileo Galilei, Via Pier Cironi 16 - 59100
PRATO, Italy 
\par\end{center}}

\begin{center}
\textit{E-mail address:} \textcolor{blue}{cordac.galilei@gmail.com}
\par\end{center}
\begin{abstract}
Realizing a gravitational wave (GW) astronomy in next years is a great
challenge for the scientific community. By giving a significant amount
of new information, GWs will be a cornerstone for a better understanding
of gravitational physics. In this paper we re-discuss that the GW
astronomy will permit to solve a captivating issue of gravitation.
In fact, it will be the definitive test for Einstein's general relativity
(GR), or, alternatively, a strong endorsement for extended theories
of gravity (ETG). 
\end{abstract}

\section{Introduction}

The scientific community hopes in a first direct detection of GWs
in next years \cite{key-1}. The realization of a GW astronomy, by
giving a significant amount of new information, will be a cornerstone
for a better understanding of gravitational physics. In fact, the
discovery of GW emission by the compact binary system PSR1913+16,
composed by two Neutron Stars \cite{key-2}, has been, for physicists
working in this field, the ultimate thrust allowing to reach the extremely
sophisticated technology needed for investigating in this field of
research. In this paper we re-discuss that the GW astronomy will permit
to solve a captivating issue of gravitation. In fact, it will be the
definitive test for Einstein's GR, or, alternatively, a strong endorsement
for ETG \cite{key-3}. 

Although Einstein's GR \cite{key-4} achieved great success (see for
example the opinion of Landau who says that GR is, together with quantum
field theory, the best scientific theory of all \cite{key-5}) and
withstood many experimental tests, it also displayed many shortcomings
and flaws which today make theoreticians question whether it is the
definitive theory of gravity, see \cite{key-6}-\cite{key-19} and
references within. As distinct from other field theories, like the
electromagnetic theory, GR is very difficult to quantize. This fact
rules out the possibility of treating gravitation like other quantum
theories, and precludes the unification of gravity with other interactions.
At the present time, it is not possible to realize a consistent quantum
gravity theory which leads to the unification of gravitation with
the other forces. From an historical point of view, Einstein believed
that, in the path to unification of theories, quantum mechanics had
to be subjected to a more general deterministic theory, which he called
\emph{generalized theory of gravitation}, but he did not obtain the
final equations of such a theory (see for example the biography of
Einstein which has been written by Pais \cite{key-20}). At present,
this point of view is partially retrieved by some theorists, starting
from the Nobel Laureate G. \textquoteright{}t Hooft \cite{key-21}. 

One can define \textit{ETG} those semi-classical theories where the
Lagrangian is modified, in respect of the standard Einstein-Hilbert
gravitational Lagrangian \cite{key-5}, adding high-order terms in
the curvature invariants (terms like $R^{2}$, $R^{\alpha\beta}R_{\alpha\beta}$,
$R^{\alpha\beta\gamma\delta}R_{\alpha\beta\gamma\delta}$, $R\Box R$,
$R\Box^{k}R$) or terms with scalar fields non-minimally coupled to
geometry (terms like $\phi^{2}R$) \cite{key-6}-\cite{key-19}. In
general, one has to emphasize that terms like those are present in
all the approaches to the problem of unification between gravity and
other interactions. Additionally, from a cosmological point of view,
such modifications of GR generate inflationary frameworks which are
very important as they solve many problems of the standard universe
model (see \cite{key-22} for a review). 

In the general context of cosmological evidence, there are also other
considerations which suggest an extension of GR. As a matter of fact,
the accelerated expansion of the universe, which is observed today,
implies that cosmological dynamics is dominated by the so called Dark
Energy, which gives a large negative pressure. This is the standard
picture, in which this new ingredient is considered as a source on
the right-hand side of the field equations. It should be some form
of un-clustered, non-zero vacuum energy which, together with the clustered
Dark Matter, drives the global dynamics. This is the so called \textquotedblleft{}concordance
model\textquotedblright{} ($\Lambda$CDM) which gives, in agreement
with the CMBR, LSS and SNeIa data, a good picture of the observed
Universe today, but presents several shortcomings such as the well
known \textquotedblleft{}coincidence\textquotedblright{} and \textquotedblleft{}cosmological
constant\textquotedblright{} problems \cite{key-23}. An alternative
approach is changing the left-hand side of the field equations, to
see if the observed cosmic dynamics can be achieved by extending General
Relativity \cite{key-6}-\cite{key-19}. In this different context,
it is not required to find candidates for Dark Energy and Dark Matter,
that, till now, have not been found; only the \textquotedblleft{}observed\textquotedblright{}
ingredients, which are curvature and baryonic matter, have to be taken
into account. Considering this point of view, one can think that gravity
is different at various scales and there is room for alternative theories
\cite{key-6}-\cite{key-19}. In principle, the most popular Dark
Energy and Dark Matter models can be achieved considering $f(R)$
theories of gravity, where $R$ is the Ricci curvature scalar, and/or
scalar-sensor gravity (STG) \cite{key-6}-\cite{key-19}. In this
picture, even the sensitive detectors for gravitational waves (GWs),
like bars and interferometers, whose data analysis recently started
\cite{key-1}, could, in principle, be important to confirm or rule
out the physical consistency of GR or of any other theory of gravitation.
This is because, in the context of ETG, some differences between GR
and the others theories can be pointed out starting from the linearized
theory of gravity, see \cite{key-3} and \cite{key-24}-\cite{key-28}. 

Now, let us consider this issue in more detail.

\section{Using gravitational waves to discriminate}

GWs are a consequence of Einstein's GR \cite{key-4}, which presuppose
GWs to be ripples in the space-time curvature travelling at light
speed \cite{key-29,key-30}. Only asymmetric astrophysics sources
can emit GWs. The most efficient are coalescing binaries systems,
while a single rotating pulsar can rely only on spherical asymmetries,
usually very small. Supernovae could have relevant asymmetries, being
potential sources\cite{key-1}.

The most important cosmological source of GWs is, in principle, the
so called stochastic background of GWs which, together with the Cosmic
Background Radiation (CBR), would carry, if detected, a huge amount
of information on the early stages of the Universe evolution \cite{key-25},
\cite{key-31}-\cite{key-35}. The existence of a relic stochastic
background of GWs is a consequence of generals assumptions. Essentially
it derives from a mixing between basic principles of classical theories
of gravity and of quantum field theory \cite{key-31}-\cite{key-34}.
The model derives from the inflationary scenario for the early universe
\cite{key-22}, which is tuned in a good way with the WMAP data on
the CBR (in particular exponential inflation and spectral index $\approx1$
\cite{key-36}. The GWs perturbations arise from the uncertainty principle
and the spectrum of relic GWs is generated from the adiabatically-amplified
zero-point fluctuations \cite{key-31}-\cite{key-34}. The analysis
has been recently generalized to ETG in \cite{key-25} and \cite{key-35}.

In 1957, F.A.E. Pirani, who was a member of the Bondi's research group,
proposed the geodesic deviation equation as a tool for designing a
practical GW detector \cite{key-37}. Pirani showed that if a GW propagates
in a spatial region where two test masses are present, the effect
is to drive the masses to have oscillations. 

In 1959, Joseph Weber studied a detector that, in principle, might
be able to measure displacements smaller than the size of the nucleus
\cite{key-38}. He developed an experiment using a large suspended
bar of aluminium, with a high resonant $Q$ at a frequency of about
$1\mbox{ }kHz$. Then, in 1960, he tried to test the general relativistic
prediction of GWs from strong gravity collisions \cite{key-39} and,
in 1969, he claimed evidence for observation of gravitational waves
(based on coincident signals) from two bars separated by $1000\mbox{ }km$
\cite{key-40}. He also proposed the idea of doing an experiment to
detect gravitational waves using laser interferometers \cite{key-40}.
In fact, all the modern detectors can be considered like being originated
from early Weber's ideas \cite{key-1}.

In recent papers \cite{key-26,key-27} it has been shown that GWs
from ETG generate different oscillations of test masses, with respect
to GWs from standard GR. Thus, an accurate analysis of such a motion
can be used in order to discriminate among various theories. 

In general, GWs manifest them-self by exerting tidal forces on the
test-masses which are the mirror and the beam-splitter in the case
of an interferometer \cite{key-1}. 

Working with $G=1$, $c=1$ and $\hbar=1$ (natural units), the line
element for a GW arising from standard GR and propagating in the $z$
direction is \cite{key-3,key-28,key-41} 

\begin{equation}
ds^{2}=dt^{2}-dz^{2}-(1+h_{+})dx^{2}-(1-h_{+})dy^{2}-2h_{\times}dxdy,\label{eq: metrica TT totale}\end{equation}

where $h_{+}(t-z)$ and $h_{\times}(t-z)$ are the weak perturbations
due to the $+$ and the $\times$ polarizations which are expressed
in terms of synchronous coordinates in the transverse-traceless (TT)
gauge \cite{key-3,key-28,key-41}. 

In the case of standard GR the motion of test masses, due to GWs and
analysed in the gauge of the local observer, is well known \cite{key-41}.
By putting the beam-splitter in the origin of the coordinate system,
the components of the separation vector are the mirror's coordinates.
At first order in $h_{+}$, the displacements of the mirror due by
the $+$ polarization of a GW propagating in the $z$ direction are
given by \cite{key-41}:

\begin{equation}
\delta x_{M}(t)=\frac{1}{2}x_{M0}h_{+}(t)\label{eq: spostamento lungo x GR}\end{equation}

and

\begin{equation}
\delta y_{M}(t)=-\frac{1}{2}y_{M0}h_{+}(t),\label{eq: spostamento lungo y GR}\end{equation}

where $x_{M0}$ and $y_{M0}$ are the initial (unperturbed) coordinates
of the mirror. The $\times$ polarization generates an analogous motion
for test masses which are rotated of 45-degree with respect the $z$
axis \cite{key-41}.

In all ETG a third \emph{massive} polarization of GWs is present \cite{key-3},
\cite{key-24}-\cite{key-27}, which is usually labelled with $\Phi,$
and the line element for such a third polarization can be always put
in a conformally flat form in both of the cases of STG and $f(R)$
theories \cite{key-24}-\cite{key-27}: \begin{equation}
ds^{2}=[1+\Phi(t-v_{G}z)](-dt^{2}+dz^{2}+dx^{2}+dy^{2}).\label{eq: metrica puramente scalare}\end{equation}

$v_{G}$ in Eq. (\ref{eq: metrica puramente scalare}) is the particle's
velocity (the group velocity in terms of a wave-packet \cite{key-3},
\cite{key-24}-\cite{key-27}). In the case of STG the third mode
can be massless. In that case $v_{G}=1$ and, at first order in $\Phi$,
the displacements of the mirror due to these massless scalar GWs are
given by \cite{key-26}

\begin{equation}
\delta x_{M}(t)=\frac{1}{2}x_{M0}\Phi(t)\label{eq: spostamento lungo x}\end{equation}

and

\begin{equation}
\delta y_{M}(t)=\frac{1}{2}y_{M0}\Phi(t).\label{eq: spostamento lungo y}\end{equation}

In the case of massive scalar GWs and of $f(R)$ theories it is \cite{key-26,key-27}

\begin{equation}
\begin{array}{c}
\delta x_{M}(t)=\frac{1}{2}x_{M0}\Phi(t)\\
\\\delta y_{M}(t)=\frac{1}{2}y_{M0}\Phi(t)\\
\\\delta z_{M}(t)=-\frac{1}{2}m^{2}z_{M0}\psi(t),\end{array}\label{eq: spostamenti}\end{equation}

where \cite{key-26,key-27} \begin{equation}
\ddot{\psi}(t)\equiv\Phi(t).\label{eq: definizione di psi}\end{equation}

Note: the most general definition is $\psi(t-v_{G}z)+a(t-v_{G}z)+b$,
but one assumes only small variations of the positions of the test
masses, thus $a=b=0$ \cite{key-26,key-27}. Then, in the case of
massive GWs a longitudinal component is present because of the presence
of a small mass $m$ \cite{key-26,key-27}. As the interpretation
of $\Phi$ is in terms of a wave-packet, solution of the Klein-Gordon
equation \cite{key-26,key-27}

\begin{equation}
\square\Phi=m^{2}\Phi,\label{eq: KG}\end{equation}

it is also

\begin{equation}
\psi(t-v_{G}z)=-\frac{1}{\omega^{2}}\Phi(t-v_{G}z).\label{eq: psi 2}\end{equation}

Thus, if advanced projects on the detection of GWs will improve their
sensitivity allowing to perform a GWs astronomy (this is due because
signals from GWs are quite weak) \cite{key-1}, one will only have
to look which is the motion of the mirror in respect to the beam splitter
of an interferometer in the locally inertial coordinate system in
order to understand which is the correct theory of gravity. If such
a motion will be governed only by Eqs. (\ref{eq: spostamento lungo x GR})
and (\ref{eq: spostamento lungo y GR}) we will conclude that GR is
the ultimate theory of gravity. If the motion of the mirror is governed
also by Eqs. (\ref{eq: spostamento lungo x}) and (\ref{eq: spostamento lungo y}),
in addition to the motion arising from Eqs. (\ref{eq: spostamento lungo x GR})
and (\ref{eq: spostamento lungo y GR}), we will conclude that massless
STG is the correct theory of gravitation. Finally, if the motion of
the mirror is governed also by Eqs. (\ref{eq: spostamenti}) in addition
to the ordinary motion of Eqs. (\ref{eq: spostamento lungo x GR})
and (\ref{eq: spostamento lungo y GR}), we will conclude that the
correct theory of gravity will be massive STG which is equivalent
to $f(R)$ theories. Even if such signals will be quite weak, a consistent
GWs astronomy will permit to understand which is the direction of
the propagating GW by using coincidences between various detectors
and to compute a hypothetical group velocity $v_{G}$ by using delay
times, thus, all the quantities of the above equations could be, in
principle, determined.

\section{Conclusion remarks}

We re-discussed that the GW astronomy will permit to solve a captivating
issue of gravitation. If advanced projects on the detection of GWs
will improve their sensitivity allowing to perform a GWs astronomy,
such a GWs astronomy will be the definitive test for Einstein's GR,
or, alternatively, a strong endorsment for ETG. In fact, a careful
analysis of the motion of the mirror of the interferometer with respect
to the beam splitter will permit to discriminate among GR and ETG.

\subsection*{Acknowledgements }

I thank the Associazione Scientifica Galileo Galilei, for supporting
this paper. I strongly thank Emilio Elizalde and the other organizers
for inviting me at the Workshop \char`\"{}Cosmology, the Quantum Vacuum
and Zeta Functions\char`\"{} for the celebration of Emilio Elizalde's
sixtieth birthday.

\end{document}